\documentclass[twocolumn,floatfix,showpacs]{revtex4}
\usepackage{amsfonts}
\usepackage{amsmath}
\usepackage{amssymb}
\usepackage{mathrsfs}
\usepackage[english,activeacute]{babel}
\usepackage{graphicx}
\usepackage{subfigure}
\usepackage{multirow}
\usepackage{epsfig}
\usepackage{float}

 \usepackage{color}

\begin{document}

\title{{\small Zeitschrift f. Naturforschung A 73(10), pp. 883-892 (2018)
}\\ \bigskip
{\Large Traveling wave solutions for wave equations with two exponential nonlinearities}}

\author{Stefan  C.  Mancas}\email{mancass@erau.edu}
\affiliation{Department of Mathematics, Embry-Riddle Aeronautical University, Daytona Beach, FL 32114-3900, USA}

\author{Haret C. Rosu}\email{hcr@ipicyt.edu.mx (Corresponding-author)}
\affiliation{IPICyT, Instituto Potosino de Investigacion Cientifica y Tecnologica,\\
Camino a la presa San Jos\'e 2055, Col. Lomas 4a Secci\'on, 78216 San Luis Potos\'{\i}, S.L.P., Mexico}%

\author{Maximino P\'erez-Maldonado}\email{maximino.perez@ipicyt.edu.mx}
\affiliation{IPICyT, Instituto Potosino de Investigacion Cientifica y Tecnologica,\\
Camino a la presa San Jos\'e 2055, Col. Lomas 4a Secci\'on, 78216 San Luis Potos\'{\i}, S.L.P., Mexico}%
%


\begin{abstract}
We use a simple method that leads to the integrals involved in obtaining the traveling-wave solutions of wave equations with one and two exponential nonlinearities. When the constant term in the integrand is zero, implicit solutions in terms of hypergeometric functions are obtained
while when that term is nonzero, all the basic traveling-wave solutions of Liouville, Tzitz\'eica and their variants, as well as sine/sinh-Gordon equations with important applications in the phenomenology of nonlinear physics and dynamical systems are found through a detailed study of the corresponding elliptic equations.
\\

\pacs{02.30.Hq, 04.20.Jb, 02.30.Ik \hfill {\bf arXiv: 1708.00542v3}}


\end{abstract}

\maketitle

\bigskip

\section{Introduction}
Some of the best known and well-studied hyperbolic nonlinear second-order differential equations are the sine-Gordon equations \cite{sG}, its variant the sinh-Gordon equation, the Tzitz\'eica equation \cite{1a,1b,1c} and its variants, such as the Dodd-Bullough equation \cite{2} and the Dodd-Bullough-Mikhailov equation \cite{3,4}, and last but not least, the Liouville equation \cite{L53}, which is a simpler case in this class. Discovered in the realm of differential geometry of surfaces with particular properties of the curvature, like in the sine-Gordon (1862) and Tzitz\'eica (1907) cases, or during the study of such surfaces as stated by Liouville (1853) in his short note, all of them have been revived much later when it became clear that they have important applications in solid state physics, nonlinear optics, biological physics, and quantum field theory through their soliton type solutions that can describe a variety of dynamical entities. This is especially true for the sine-Gordon equation whose soliton solutions have been identified with dislocations in crystals, fluxons in long Josephson junctions, waves in magnetic materials and superfluids, nonlinear DNA and microtubule excitations, neural impulses, and muscular contractions, among others \cite{B71,II13}.

In this article, we will approach all these equations as particular cases of second-order differential equations with two exponential nonlinear terms of the form
\begin{equation}\label{eq2}
\psi_{uv}=\alpha e^{a\psi}+\beta e^{b\psi}~,
\end{equation}
where $a$ and $b$ are nonzero real constants, while $\alpha$ and $\beta$ are real constants not simultaneously zero.
The main advantage of this approach is to have a unifying treatment of these famous equations, which in general are considered separately by the majority of authors, as illustrative examples of their solution methods.
In the 1970s, during the remarkable advance in the solution method for nonlinear evolution equations brought by the inverse scattering method, Dodd and Bullough \cite{db76} posed and solved the problem of which equations of the form $y_{xt}=f(y)$ admit infinitely many integrals of motion, a property of soliton evolution equations discovered by Zakharov and Shabat in their breakthrough paper of 1972 \cite{zs72}. Dodd and Bullough showed that, beyond the linear case, the only allowed hyperbolic nonlinear equations with this property are precisely of the Liouville, sine/sinh-Gordon, and Tzitz\'eica forms and the variants of the latter. This was a confirmation of the fact that these type of equations have soliton solutions, some of which were already known at that time.

On the other hand, this kind of equations can be turned into polynomial nonlinear equations
\begin{equation}\label{eq1}
\frac{\partial^2}{\partial u \partial v}\log h=\alpha h^a+\beta h^b,
\end{equation}
by using the change of variables $\psi=\log h$.

\medskip

Along the two characteristics  $z=u-\lambda v$, $t=u+\lambda v$, $\lambda \ne 0$, (\ref{eq2}) becomes the nonlinear wave equation
\begin{equation}\label{eq3}
\psi_{tt}-\psi_{zz}=\frac 1 \lambda \left( \alpha e^{a\psi}+\beta e^{b\psi} \right)
\end{equation}
or in the logarithmic variable
\begin{equation}\label{eq4}
h(h_{tt}-h_{zz})-({h_t}^2-{h_z}^2)=\frac{h^2}{\lambda}\left(\alpha h^a+\beta h^b\right).
\end{equation}

 Furthermore,  the usage of the traveling-wave ansatz $h(z,t)=h(\xi)$ with $\xi=k z- \omega t$, and $k \ne \pm \omega $ in (\ref{eq4}) yields the following ordinary differential equation (ODE)
\begin{equation}\label{eq5}
hh_{\xi\xi}-{h_\xi}^2=\frac{h^2}{\lambda \gamma}\left(\alpha h^a+\beta h^b\right)\equiv f(h),
\end{equation}
with  $\gamma =\omega^2-k^2\ne 0$.
Once the traveling variable reduction is performed, one cannot avoid to recall that there is a multitude of papers on a variety of effective methods
to solve the resulting ODE's based on polynomial ansatze of the solution, in general stemming from the breakthrough tanh-method of Malfliet and Hereman \cite{mh1,m2,pd}. We mention the equivalent $G'/G$-expansion method \cite{K}, the sinh-cosh \cite{WW2} and tanh-coth methods \cite{sak1}, the $Q$-function method \cite{KQ,sak2}, and the more powerful Jacobi elliptic function method and their extended versions which can be used for more complicated equations such as the double sine-Gordon \cite{fz} or to derive doubly periodic wave solutions of a variety of Boussinesq-like equations
\cite{sak1}. In fact, the $G'/G$-expansion method and the tanh method have been already applied to Tzitz\'eica's equation and its variants in \cite{4,aba}.
However, for the ODE in (\ref{eq5}), we will apply a simple trick that we used in a previous paper \cite{5} to reduce it to a Bernoulli equation, thus allowing us to obtain easily all the basic solutions both in its full generality and simplified to the important cases mentioned above. This reduction is apparently not well known in this context which motivated us to write the present paper.

\section{The implicit solution}
A simple method to solve ODEs of type (\ref{eq5}) is to let $h_\xi=u(h(\xi))$, and use $h_{\xi\xi}= u\frac{du}{dh}$ \cite{5} to obtain
\begin{equation}\label{eq5a}
h u\frac{du}{dh}-u^2=f(h)~,
\end{equation}
 which can be turned into a Bernoulli equation using the substitution $u^2=z$
\begin{equation}\label{eq5b}
\frac{dz}{dh}-\frac 2 h z=\frac{2}{h}f(h)~.
\end{equation}
The solution for this equation is
\begin{equation}\label{eq5c}
z=h^2\left(c_0+2 \int \frac{f(h)}{h^3}dh\right)~,
\end{equation}
and using back the transformations $u=\pm \sqrt z =\frac{dh}{d\xi}$, we can obtain $h$ by the quadrature
\begin{equation}\label{eq5d}
\int \frac{dh}{h\sqrt{c_1+\frac{\alpha}{a} h^{a}+\frac{\beta}{b} h^{b}}}=\pm\sqrt{\frac{2}{\lambda \gamma}} \int d\xi~.
\end{equation}
In general, this quadrature can be performed only if the function $h$ satisfies the elliptic equation or if $c_1=0$. In the latter case,
one obtains the following implicit solution involving the hypergeometric function
{\small \begin{align}\label{c1zero}
h^{-b}\sqrt{\frac{\alpha  h^{a}}{a}+\frac{\beta h^{b}}{b}} \,
 _{2}F_1\left(1,1-\frac{a}{2(a-b)}; 1-\frac{b}{2(a-b)};-\frac{b\alpha}{a\beta}h^{a-b}\right) \nonumber \\
 =\mp \frac{\beta}{\sqrt{2 \lambda \gamma}} (\xi-\xi_0)~,
\end{align}}
which also implies $\beta\neq 0$.
We will show in the next section how the quadrature in (\ref{eq5d}) is solved in the important particular cases mentioned in the introduction for $c_1\neq 0$.

\section{Particular cases}
\subsection{Liouville equation}
We start with the simplest case, which is the Liouville equation, 
`the widely known example of an exactly integrable nonlinear partial differential equation' \cite{23} in mathematical physics. For example, in cosmology, the inflationary expansion epoch of the early universe is usually generated by means of one or several scalar fields with an exponential potential and are dubbed Liouville cosmologies \cite{a15}.

\begin{figure}[ht!] 
\centering 
\includegraphics[width=0.45\textwidth]{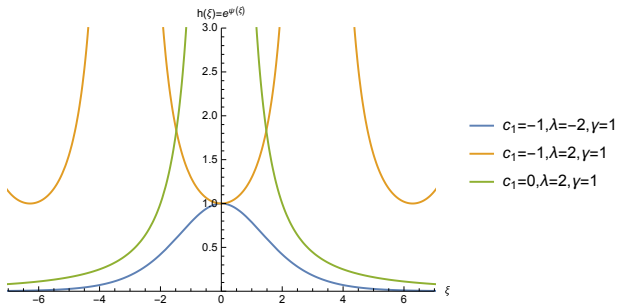}
\caption{The soliton and periodic solutions (\ref{eq17bis}) and the rational solution (\ref{eq17tris}) of the Liouville equation.}
\label{m1}
\end{figure}

The Liouville equation corresponds to $\alpha=1,~\beta=0,~a=1,~b=0$, which is
\begin{equation}\label{eq14}
\frac{\partial^2}{\partial u \partial v}\log h= h~.
\end{equation}
The quadrature in~(\ref{eq5d}) takes the form
\begin{equation}\label{eq15}
\int \frac{dh}{h\sqrt{c_1+ h}}=\pm\sqrt{\frac{2}{\lambda \gamma}} (\xi-\xi_0)~,
\end{equation}
and the function $h$ is obtained by solving the elliptic equation
\begin{equation}\label{eq6}
{h_\xi}^2=a_3 h^3+a_2h^2+a_1 h+a_0
\end{equation}
with the coefficients given by the system
\begin{equation}\label{eq16}
\begin{array}{l}
a_0=0~,\\
a_1=0~,\\
a_2=\frac{2c_1}{\lambda \gamma}~,\\
a_3=\frac{2}{\lambda \gamma}~.
\end{array}
\end{equation}
For convenience, we denote $r=\frac{1}{\lambda \gamma}$ and $p=\frac{2c_1}{\lambda \gamma}$; then~(\ref{eq6}) becomes
the reduced elliptic equation
\begin{equation}\label{eq16a}
h_\xi^2=2rh^3+p h^2~,
\end{equation}
with soliton solution if $p>0$, or periodic solution if $p<0$
\begin{equation}\label{eq16b}
\begin{array}{l}
h(\xi)=-\frac{p}{2r}\mathrm{sech}^2\left[\frac 1 2 \sqrt{p}(\xi-\xi_0)\right]~, \quad p>0\\
h(\xi)=-\frac{p}{2r}\mathrm{sec}^2\left[\frac 1 2 \sqrt{-p}(\xi-\xi_0)\right]~, \quad p<0~,
\end{array}
\end{equation}
and by using system (\ref{eq16}) give the solutions
\begin{equation}\label{eq17bis}
\begin{array}{l}
h(\xi)=-c_1 \mathrm{sech}^2\left(\sqrt{\frac{ c_1}{2\lambda \gamma}}(\xi-\xi_0)\right)~, \quad \frac{c_1}{2\lambda\gamma}>0\\
h(\xi)=-c_1 \mathrm{sec}^2\left(\sqrt{\frac{-c_1}{2\lambda \gamma}}(\xi-\xi_0)\right)~, \quad \frac{c_1}{2\lambda\gamma}<0~.
\end{array}
\end{equation}
Since we are in the case $\beta=0$, we cannot apply (\ref{c1zero}) when $c_1=0$. However, the integration of (\ref{eq15}) is easily performed and corresponds to the most degenerate case of the Weierstrass elliptic equation (\ref{eq8}) when both germs $g_2$ and $g_3$ are zero, which  leads to the rational solution
\begin{equation}\label{eq17tris}
h(\xi)=\frac{2\lambda\gamma}{(\xi-\xi_0)^2}~.
\end{equation}
Plots of all these three types of Liouville solutions are given in Figure~(\ref{m1}). The only nonsingular solution is the soliton one, which is the usual solution employed in the literature.

\subsection{The Tzitz\'eica equation}
Tzitz\'eica's equation,
\begin{equation}\label{eq5e}
\frac{\partial^2}{\partial u \partial v}\log h=h-\frac{1}{h^2}~,
\end{equation}
emerged in 1907-1910 in the area of geometry, but only after 80 years has it been found to have applications in physics. For example, Euler's equations for an ideal gas with a special equation of motion can be reduced to the Tzitz\'eica equation, and a 2+1-dimensional system in magnetohydrodynamics has been shown to be in one-to-one correspondence with it \cite{t-app1,t-app2}. Very recently, dark optical solitons and traveling waves of the Tzitz\'eica type have been also discussed in the literature \cite{t-app4,t-app5}.

For Tzitz\'eica's equation, we identify the constants in (\ref{eq2}) as $\alpha=1,~\beta=-1,~a=1,~ b=-2$, which gives the quadrature
\begin{equation}\label{eq5f}
\int \frac{dh}{\sqrt{h^3+c_1h^2+\frac 1 2}}=\pm\sqrt{\frac{2}{\lambda \gamma}} (\xi-\xi_0)~.
\end{equation}
For $c_1=0$, the quadrature reads
\begin{equation}\label{eq5f1}
\int \frac{dh}{\sqrt{h^3+\frac 1 2}}=\pm\sqrt{\frac{2}{\lambda \gamma}} (\xi-\xi_0)~,
\end{equation}
with solution given by the equianharmonic function of case 2 (i) below, while
implicitly, the equianharmonic $h$ satisfies~(\ref{c1zero}) and simplifies to
\begin{equation}\label{tzi}
h\, _2F_1\left(\frac 1 2,\frac 1 3; \frac 4 3 ;-2h^3\right)=\pm \frac{\xi-\xi_0}{\sqrt{\lambda \gamma}}~.
\end{equation}

The solution $h$ is obtained explicitly by solving the elliptic equation (\ref{eq6})
with coefficients given by the system
\begin{equation}\label{eq7}
\begin{array}{l}
a_0=
r~,\\
a_1=0~,\\
a_2=
2c_1r~,\\
a_3=
2r~,
\end{array}
\end{equation}
which becomes
\begin{equation}\label{eq6bis}
{h_\xi}^2=2 r h^3+ph^2+r~.
\end{equation}
Using the  scale shift transformation
\begin{equation}\label{tra}
h(\xi)=\frac1 r \left(2 \wp(\xi; g_2,g_3)-\frac{p}{6}\right)~,
\end{equation}
equation~(\ref{eq6bis}) becomes the Weierstrass equation
\begin{equation}\label{eq8}
\wp_{\xi}^{\,\,2}=4 \wp^3-g_2 \wp -g _3~.
\end{equation}
The germs of the Weierstrass function are given by
\begin{equation}\label{eq9}
\begin{array}{l}
g_2=\frac{a_2^2-3 a_1a_3}{12}=\frac{p^2}{12}=2(e_1^2+e_2^2+e_3^2)~,\\
g_3=\frac{9 a_1a_2a_3-27 a_0 a_3^2-2 a_2^3}{432}=-\frac{1}{4}\left(r^3+\frac{p^3}{54}\right)=4e_1e_2e_3
\end{array}
\end{equation}
and together with the modular discriminant
\begin{align}\label{eq10}
&\Delta={g_{2}}^3-27 {g_{3}}^2=-\frac{r^3}{16}(p^3+27 r^3) \nonumber\\
&=16(e_1-e_2)^2(e_1-e_3)^2(e_2-e_3)^2
\end{align}
are used to classify the solutions of~(\ref{eq6bis}), where the constants $e_i$ are the roots of the cubic polynomial
\begin{equation}
s_3(t)=4t^3-g_2t-g_3=4(t-e_1)(t-e_2)(t-e_3)=0~.
\end{equation}

\begin{description}
	\item [Case 1.]
If $\Delta  \equiv 0 \Rightarrow p=-3 r\Rightarrow c_1=-\frac 3 2$. This degenerate case implies that $s_3(t)$ has repeated root of multiplicity two.
Then the  Weierstrass solutions can be simplified since $\wp$ degenerates into hyperbolic or trigonometric functions. Because of the degeneracy,~(\ref{eq5f}) can be  factored as
\begin{equation}\label{eq10a}
\int \frac{dh}{\sqrt{(h-1)^2\left(h+\frac 1 2\right)}}=\pm\sqrt{\frac{2}{\lambda \gamma}} (\xi-\xi_0)~.
\end{equation}
Depending on the sign of $g_3$, we have the following sub-cases:
		\begin{enumerate}
		 \item [Case (1a).]$r>0$ with  $g_2>0 \Rightarrow g_3=-\frac{r^3}{8}<0 \Rightarrow \lambda\gamma>0$, so~(\ref{eq10a}) has the dark soliton solution
\begin{equation} \label{eq11}
h(\xi)=
1-\frac{3}{2}\mathrm{sech}^2 \left(\frac {1}{2}\sqrt{\frac{3}{\lambda \gamma}}(\xi-\xi_0) \right)~.
\end{equation}
By  letting $e_1=e_2=\hat e>0$, then $e_3=-2\hat e<0$, hence
\begin{equation}\label{eq120}
\begin{array}{l}
	g_2=12{\hat e}^2>0~,\\
	g_3=-8{\hat e}^3<0\\
\end{array}
\end{equation}
and the Weierstrass $\wp$ solution to~(\ref{eq8}) reduces to
\begin{equation} \label{eq130}
\wp(\xi;12{\hat e}^2,-8{\hat e}^3)=\hat e+3\hat e~ \mathrm{csch}^2(\sqrt{3\hat e}\xi)~.
\end{equation}

For $\hat e=\frac{1}{4\lambda \gamma}>0$, the Weierstrass solution replaced by (\ref{eq130}) gives the singular (blow-up) soliton
\begin{equation} \label{eq101}
h(\xi)=1+\frac 32\mathrm{csch}^2 \left(\frac {1}{2}\sqrt{\frac{3}{\lambda \gamma}}(\xi-\xi_0) \right)~.
\end{equation}
		\item [Case (1b).] $r<0$ with  $g_2>0 \Rightarrow g_3=-\frac{r^3}{8}>0 \Rightarrow \lambda\gamma<0$,~(\ref{eq10a}) has the following solution with periodic negative singularities:
\begin{equation} \label{eq12}
h(\xi)=
1-\frac{3}{2}\mathrm{sec}^2 \left(\frac {1}{2}\sqrt{\frac{3}{-\lambda \gamma}}(\xi-\xi_0) \right)~.
\end{equation}
		\end{enumerate}
		 By  letting $e_2=e_3=-\tilde {e}<0$ with $\tilde {e}>0$, then $e_1=2 \tilde{e}>0$, and hence
\begin{equation}\label{eq140}
\begin{array}{l}
	g_2=12{\tilde {e}}^2>0\\
	g_3 =8{\tilde {e}}^3>0~.\\
\end{array}
\end{equation}
 The Weierstrass $\wp$ solution reduces to
\begin{equation} \label{eq150}
\wp(\xi;12{\tilde {e}}^2,8{\tilde {e}}^3)=-{\tilde {e}}+3{\tilde {e}}~ \mathrm{csc}^2(\sqrt{3{\tilde {e}}}\xi)~.
\end{equation}
		
		For $\tilde e=-\frac{1}{4\lambda \gamma}>0$, the Weierstrass solution replaced by (\ref{eq150}) gives
\begin{equation} \label{eq1010}
h(\xi)=1-\frac 32\mathrm{csc}^2 \left(\frac {1}{2}\sqrt{\frac{3}{-\lambda \gamma}}(\xi-\xi_0) \right)~,
\end{equation}
which is similar to (\ref{eq12}). All the Tzitz\'eica solutions corresponding to these cases are displayed in Figure~(\ref{m2}). The dark soliton solution may have physical applications, especially in optics and hydrodynamics, while the other singular solutions look unphysical for the time being. However, we notice that there are already detailed mathematical studies of the blow up problem in the Tzitz\'eica case \cite{blowup}.

\begin{figure}[ht!] 
\centering 
\includegraphics[width=0.45\textwidth]{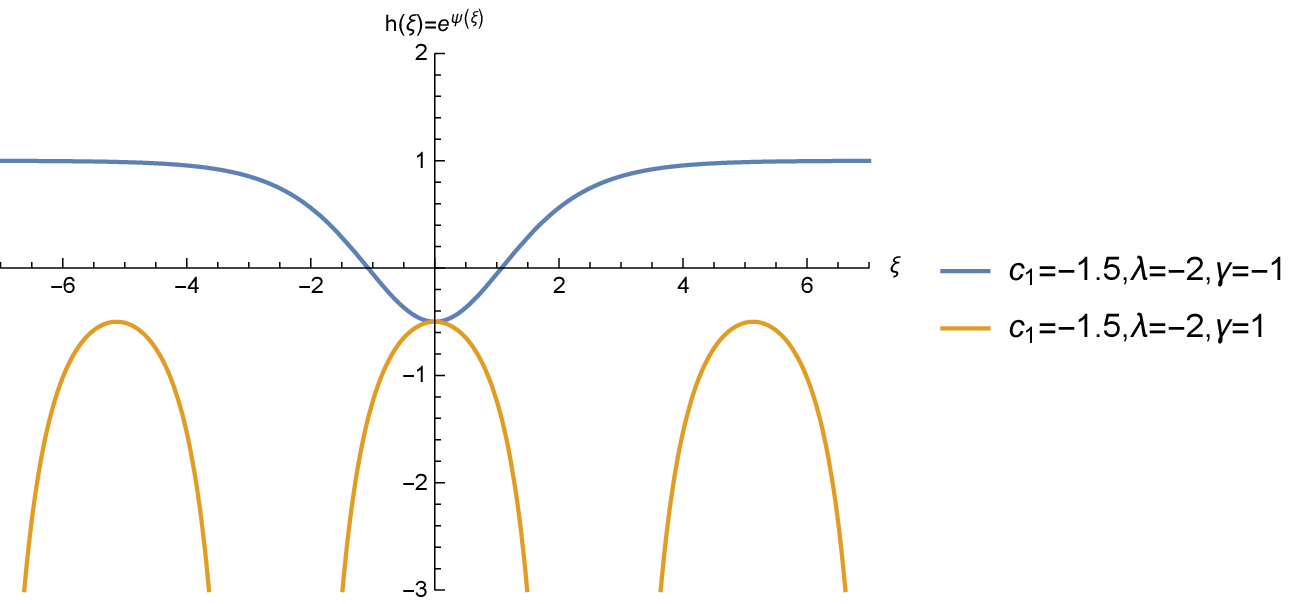}\\
\includegraphics[width=0.45\textwidth]{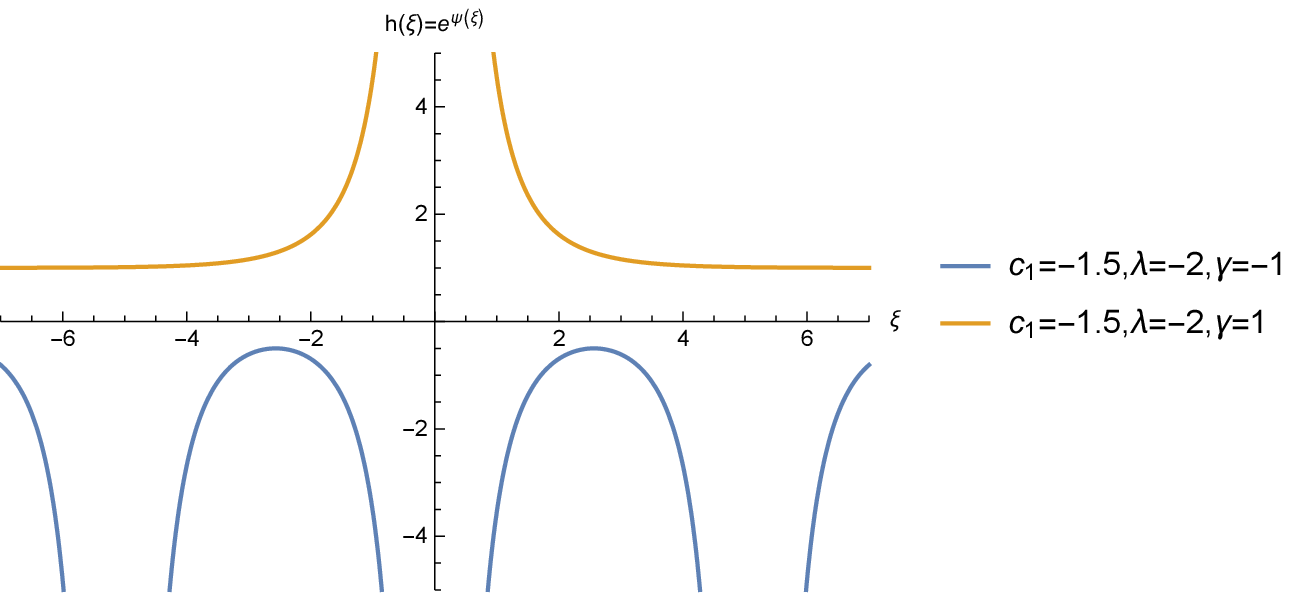}\\
\caption{The Tzitz\'eica dark soliton (\ref{eq11}) and the periodic singular solution (\ref{eq12}) (top). The Tzitz\'eica solution (\ref{eq101}), which can be also called Tzitz\'eica's singular soliton, and the periodic singular solution (\ref{eq1010}) (bottom).}
\label{m2}
\end{figure}
		
\item [Case (2).]
If $\Delta  \ne 0 \Rightarrow p \ne -3r \Rightarrow c_1 \ne -\frac 3 2$,  we include   two  particular solutions which will
fix the integration constant $c_1$ as follows:  the equianharmonic ($g_2=0$) and lemniscatic case ($g_3=0$), respectively.

{\bf i)} For the equianharmonic case, $g_2=0 \Rightarrow p=0\Rightarrow g_3=-\frac{r^3}{4}$. Because $\Delta=-\frac{27}{16}r^6<0$,
 $s_3(t)$ has a pair of conjugate complex roots, and since $c_1=0$, the  solution to~(\ref{eq5f}) reduces to
\begin{equation}\label{soli1}
h(\xi)=2 \lambda \gamma \wp\left(\xi-\xi_0;0,-\frac{1}{4 \lambda^3 \gamma^3}\right)~.
\end{equation}

{\bf ii)} For the lemniscatic case, $g_3=0  \Rightarrow p=-3 \sqrt[3]{2}r \Rightarrow g_2 =\frac{3 \sqrt[3]{4}}{4}r^2$.  Because $\Delta=\frac{27}{16}r^6>0$, $s_3(t)$ has three distinct real roots given by $e_3=-\frac{\sqrt {g_2}}{2}$, $e_2=0$, and $e_1=\frac{\sqrt {g_2}}{2}$. Although the Weierstrass unbounded function  has poles aligned on the real axis of the $\xi-\xi_0$ complex plane, we can choose $\xi_0$ in such a way to shift these poles a half of period above the real axis, so that the elliptic function simplifies using the formula \cite{Whitt}
\begin{equation}\label{sol5}
\wp(\xi;g_2,0)=e_3+(e_2-e_3)\mathrm{sn}^2\left(\sqrt{e_1-e_3}(\xi-\xi_0\rq{});m\right)
\end{equation}
with elliptic modulus $m=\sqrt{\frac{e_2-e_3}{e_1-e_3}}$.
Using the values of the roots together with $\xi_0\rq{}=0$, we obtain
\begin{equation}\label{sol6}
\wp(\xi;g_2,0)=-\frac{\sqrt{g_2}}{2}\mathrm{cn}^2\left(\sqrt[4]{g_2}\xi;\frac{\sqrt{2}}{2}\right)~.
\end{equation}
Because $c_1=-\frac{3}{\sqrt[3]{4}}$,  the solutions for the lemniscatic case are reduced using the transformation (\ref{tra}) to periodic  cnoidal waves, and they become
\begin{equation}\label{soli2}
h(\xi)=\frac{1}{\sqrt[3]{4}}\left[1-\sqrt 3~ \mathrm{cn}^2\left(\frac{\sqrt[4]{3}}{\sqrt[3]{2}\sqrt{\lambda \gamma}}\xi;\frac{\sqrt{2}}{2}\right)\right]~.
\end{equation}

\begin{figure}[ht!] 
\centering 
\includegraphics[width=0.45\textwidth]{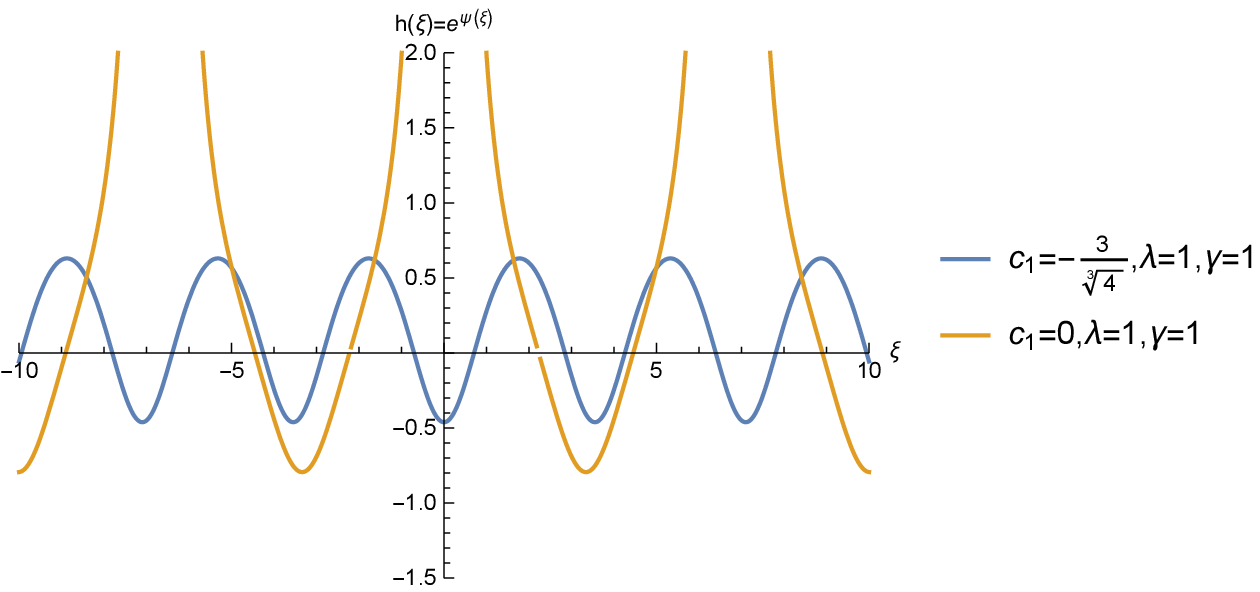}\\
\includegraphics[width=0.45\textwidth]{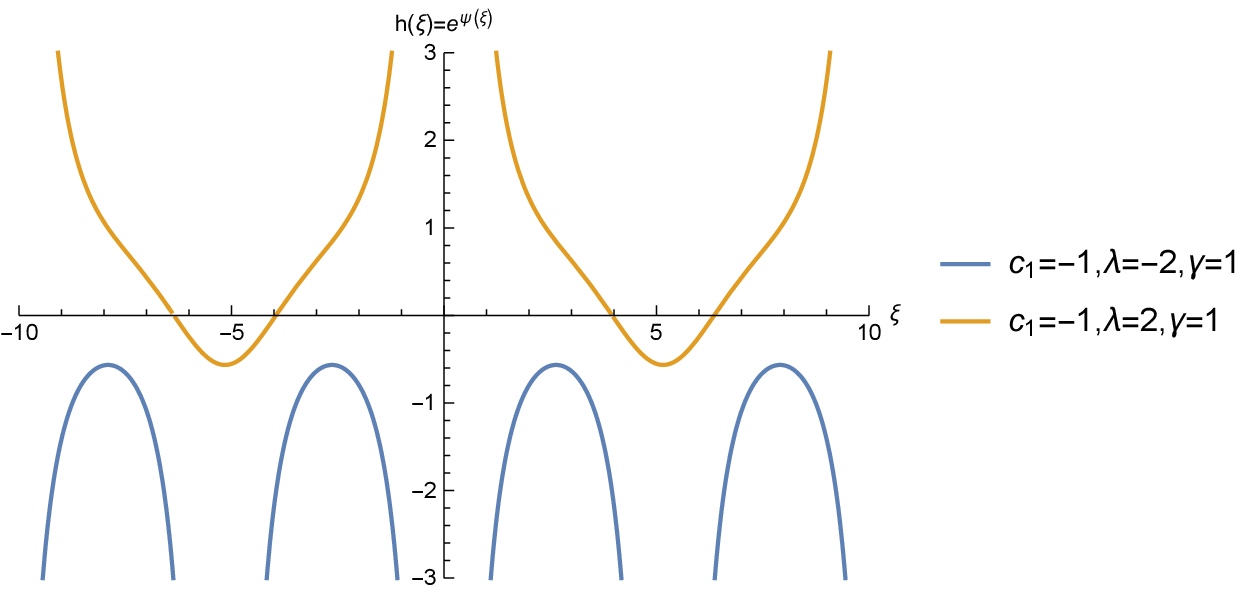}
\caption{The lemniscatic (cnoidal) and equianharmonic solutions, (\ref{soli2}) and (\ref{soli1}), respectively, of the Tzitz\'eica equation (top). The Weierstrass solution (\ref{eq13}) of the Tzitz\'eica equation (bottom).}
\label{m3}
\end{figure}

{\bf iii)} For the most general case, $g_2 \ne0 $, $g_3\ne0$, and the general solution to~(\ref{eq5f}) is
\begin{equation}\label{eq13}
h(\xi)=\lambda \gamma \left[2 \wp\left(\xi-\xi_0;\frac{{c_1}^2}{3 \lambda ^2\gamma^2},-\frac{4{c_1}^3+27}{108 \lambda^3\gamma^3}\right)-\frac{c_1}{3\lambda \gamma}\right]~.
\end{equation}
\end{description}
The equianharmonic, lemniscatic, and Weierstrass solutions of Tzitz\'eica's equation are displayed in Figure~(\ref{m3}). We notice that the only regular solutions are the periodic cnoidal ones corresponding to the lemniscatic case. All the other solutions have periodic positive or negative blow-ups and are interesting rather from the mathematical standpoint \cite{blowup} than for applications to the physical phenomenology.

\subsection{The Dodd-Bullough equation}
This variant of Tzitz\'eica's equation was introduced in the first dedicated study of the polynomial conserved quantities of the sine-Gordon equation \cite{2}. Its form in the $h$ variable is
\begin{equation}\label{eq26}
\frac{\partial^2}{\partial u \partial v}\log h=-h+\frac{1}{h^2}~,
\end{equation}
so we identify the constants to be $\alpha=-1,~\beta=1,~a=1,~ b=-2$, which gives the quadrature
\begin{equation}\label{eq27}
\int \frac{dh}{\sqrt{-h^3+c_1h^2-\frac 1 2}}=\pm\sqrt{\frac{2}{\lambda \gamma}} (\xi-\xi_0)~.
\end{equation}

In implicit form, the $h$ function satisfies~(\ref{c1zero}), which simplifies to
\begin{equation}\label{bul}
h \, _2F_1\left(\frac 1 2,\frac 1 3; \frac 4 3 ;-2h^3\right)=\mp \frac{\xi-\xi_0}{\sqrt{-\lambda \gamma}}
\end{equation}
and up to a sign, the implicit solution is the same as the Tzitzeica solution.

Using Tzitz\'eica solutions, provided that $r\rightarrow -r \Rightarrow \lambda \gamma \rightarrow - \lambda \gamma $ and $c_1 \rightarrow -c_1$, we have the cases:
\begin{description}
	\item [Case (1).]
If $\Delta  \equiv 0 \Rightarrow c_1=\frac 3 2$, then
		\begin{enumerate}
		 \item [Case (1a).]$\lambda \gamma<0$, so~(\ref{eq27}) has the soliton solution
\begin{equation} \label{eq28}
h(\xi)=
1-\frac{3}{2}\mathrm{sech}^2 \left(\frac {1}{2}\sqrt{\frac{3}{-\lambda \gamma}}(\xi-\xi_0) \right)~.
\end{equation}
The solution corresponding to (\ref{eq101}) is
\begin{equation} \label{es101}
h(\xi)=1+\frac 32\mathrm{csch}^2 \left(\frac {1}{2}\sqrt{\frac{3}{-\lambda \gamma}}(\xi-\xi_0) \right)~.
\end{equation}

		\item [Case (1b).] $\lambda \gamma>0$, so~(\ref{eq27}) has the periodic solution
\begin{equation} \label{eq29}
h(\xi)=
1-\frac{3}{2}\mathrm{sec}^2 \left(\frac {1}{2}\sqrt{\frac{3}{\lambda \gamma}}(\xi-\xi_0) \right)
\end{equation}
and the periodic solution corresponding to (\ref{eq1010}) is
\begin{equation} \label{es1010}
h(\xi)=1-\frac 32\mathrm{csc}^2 \left(\frac {1}{2}\sqrt{\frac{3}{\lambda \gamma}}(\xi-\xi_0) \right)~.
\end{equation}
		\end{enumerate}
\item [Case (2).]
If $\Delta  \ne 0   \Rightarrow c_1 \ne  \frac 3 2$, then the Weierstrass  solution of~(\ref{eq27}) is
\begin{equation}\label{eq30}
h(\xi)=-\lambda \gamma \left[2 \wp\left(\xi-\xi_0;\frac{{c_1}^2}{3 \lambda ^2\gamma^2},-\frac{4{c_1}^3-27}{108 \lambda^3\gamma^3}\right)-\frac{c_1}{3\lambda \gamma}\right]~.
\end{equation}
\end{description}

{\bf i)} For the equianharmonic case, $c_1=0$,  we have
\begin{equation}\label{soli3}
h(\xi)=-2 \lambda \gamma \wp\left(\xi-\xi_0;0, \frac{1}{4 \lambda^3 \gamma^3}\right)~.
\end{equation}

{\bf ii)} For the lemniscatic case, $c_1=\frac{3}{\sqrt[3]{4}}$,  we have
\begin{equation}\label{soli4}
h(\xi)=\frac{1}{\sqrt[3]{4}}\left[1-\sqrt 3~ \mathrm{cn}^2\left(\frac{\sqrt[4]{3}}{\sqrt[3]{2}\sqrt{-\lambda \gamma}}\xi;\frac{\sqrt{2}}{2}\right)\right]~.
\end{equation}

Respecting the rules of changing the signs of the parameters, the Dodd-Bullough solutions are identical to the Tzitz\'eica solutions
and consequently we will not plot them here.

\subsection{The Tzitz\'eica-Dodd-Bullough equation}
This variant equation reads
\begin{equation}\label{eq31}
\frac{\partial^2}{\partial u \partial v}\log h=h+\frac{1}{h^2}~,
\end{equation}
so we identify the constants to be $\alpha=1,~\beta=1,~a=1,~ b=-2$, which gives the quadrature
\begin{equation}\label{eq32}
\int \frac{dh}{\sqrt{h^3+c_1h^2-\frac 1 2}}=\mp\sqrt{\frac{2}{\lambda \gamma}} (\xi-\xi_0)~.
\end{equation}
We can use the Dodd-Bullough solutions,~(\ref{eq28})-(\ref{eq30}), with $h\rightarrow -h$ and $\xi\rightarrow -\xi$.

\subsection{The Dodd-Bullough-Mikhailov equation}
The last Tzitz\'eica variant equation reads
\begin{equation}\label{eq31}
\frac{\partial^2}{\partial u \partial v}\log h=-h-\frac{1}{h^2}~,
\end{equation}
so the constants are identified as $\alpha=-1,~\beta=-1,~a=1,~ b=-2$, which gives the quadrature
\begin{equation}\label{eq32}
\int \frac{dh}{\sqrt{-h^3+c_1h^2+\frac 1 2}}=\mp\sqrt{\frac{2}{\lambda \gamma}} (\xi-\xi_0)~.
\end{equation}
From the polynomial in the integrand, one can see that the solutions of this variant equation are obtained from Tzitz\'eica solutions by $h\rightarrow -h$ and $\xi\rightarrow -\xi$.

\subsection{The sine-Gordon equation}
According to \cite{boldin}, Tzitz\'eica's equation is the ``nearest relative" of the well-known sine-Gordon equation, which can be written as
\begin{equation}\label{eq33}
\frac{\partial^2}{\partial u \partial v}\log h=\frac{1}{2i}\left(h^i-h^{-i}\right)=\sin(\log h)~.
\end{equation}
Thus, we identify the constants to be $\alpha=\frac{1}{2i},~\beta=-\frac{1}{2i},~a=i,~ b=-i$, which gives the quadrature
\begin{equation}\label{eq34}
\int \frac{dh}{h\sqrt{c_1-\cos (\log h)}}=\pm\sqrt{\frac{2}{\lambda \gamma}} (\xi-\xi_0)~,
\end{equation}
equivalent to
\begin{equation}\label{eq35}
\int \frac{d \psi}{\sqrt{c_1-\cos (\psi)}}=\pm\sqrt{\frac{2}{\lambda \gamma}} (\xi-\xi_0)~.
\end{equation}
In implicit form, $h$ satisfies~(\ref{c1zero}), which simplifies to
\begin{equation}\label{sinhg}
h^i \sqrt{\cos (\log h)}~_2F_1\left(1,\frac 34; \frac 5 4;-h^{2 i}\right)=\mp \frac{\xi-\xi_0}{2\sqrt{2\lambda \gamma}}~.
\end{equation}
The explicit solution satisfies~\eqref{eq35}, with $c_1=0$
\begin{equation}\label{eq350}
\int \frac{d \psi}{\sqrt{\cos (\psi)}}=\mp\sqrt{\frac{2}{-\lambda \gamma}} (\xi-\xi_0)~.
\end{equation}
Using the definition of  the elliptic integral of the first kind $\xi=F(\phi;m)=\int_0^\phi\frac{d\psi}{\sqrt{1-m \sin^2{\psi}}}$ \cite{rob,mond}, (\ref{eq350}) becomes
\begin{equation}\label{eq351}
F\left(\frac{\psi}{2};2\right)=\mp\frac 1 2 \sqrt{\frac{2}{-\lambda \gamma}} (\xi-\xi_0)~.
\end{equation}
By inverting, the explicit solution is
\begin{equation}\label{eq352}
\psi(\xi)=2~ \mathrm{am}\left(\mp\frac 1 2 \sqrt{\frac{2}{-\lambda \gamma}} (\xi-\xi_0);2\right)~,
\end{equation}
where the function $\mathrm{am}(\xi;m)$ is the Jacobi amplitude function, which can also be obtained from (\ref{eq40}) for $c_1=0$.

\begin{figure}[ht!] 
\centering 
\includegraphics[width=0.45\textwidth]{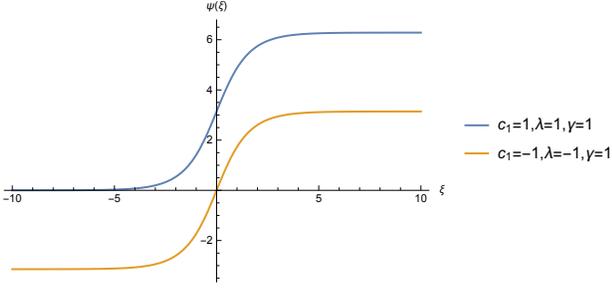} 
\caption{The arctan kink solutions (\ref{eq37}) and (\ref{eq370}) of the sine-Gordon equation.}
\label{m6}
\end{figure}

For the special case of $c_1=1$,~(\ref{eq35}) simplifies to
\begin{equation}\label{eq36}
\int \frac{d \psi}{\sin \frac{\psi}{2}}=\pm \frac{2}{\sqrt{\lambda \gamma}} (\xi-\xi_0)~.
\end{equation}
For $\lambda \gamma>0$, and using the identity $\tan \frac \theta 2=\frac{1-\cos \theta}{\sin \theta}$,
we obtain  the kink-antikink solutions
\begin{equation}\label{eq37}
\psi(\xi)=4 \arctan \left(e^{\pm\frac{\xi-\xi_0}{\sqrt{\lambda \gamma}}}\right)~.
\end{equation}
When $c_1=-1$,~(\ref{eq35}) simplifies to
\begin{equation}\label{eq360}
\int \frac{d \psi}{\cos \frac{\psi}{2}}=\pm \frac{2}{\sqrt{-\lambda \gamma}} (\xi-\xi_0)
\end{equation}
for $\lambda \gamma<0$, and using the identity $\tan\left(\frac \theta2 +\frac \pi 4\right)=\frac{1-\cos \theta}{\sin \theta}$
the shifted kink-antikink solution is
\begin{equation}\label{eq370}
\psi(\xi)=-\pi  + 4 \arctan \left(e^{\pm\frac{\xi-\xi_0}{\sqrt{-\lambda \gamma}}}\right)~.
\end{equation}
Plots of the $c_1=\pm 1$ arctan solutions (\ref{eq37}) and (\ref{eq370}) are displayed in Figure~(\ref{m6}). As is well known, these kink solutions are overwhelmingly encountered in applications \cite{B71}.

\begin{figure}[ht!] 
\centering 
\includegraphics[width=0.45\textwidth]{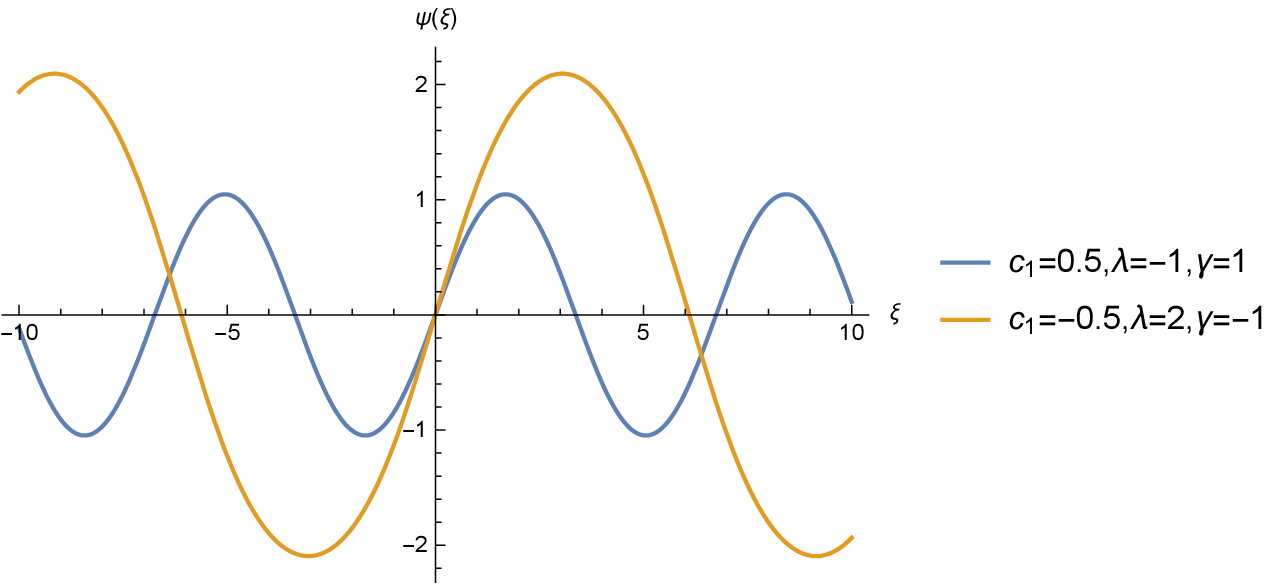}\\
\includegraphics[width=0.45\textwidth]{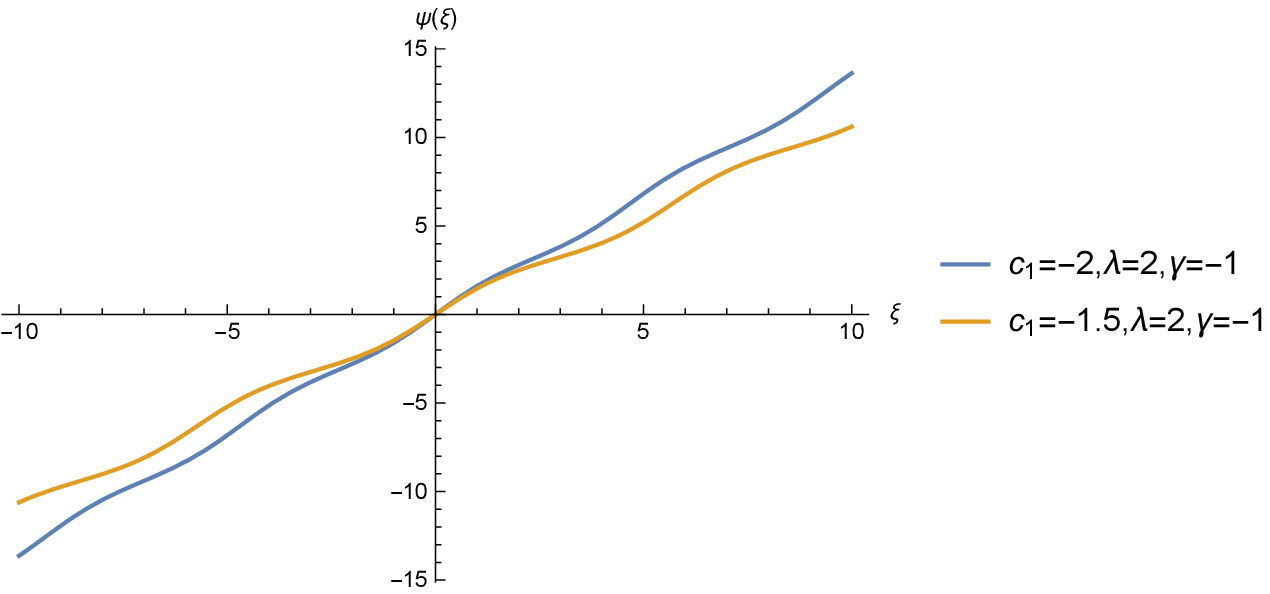}
\caption{The Jacobi amplitude solution (\ref{eq40}) of the sine-Gordon equation for $|m|>1$ (top) and $|m|<1$ (bottom).}
\label{m7}
\end{figure}

When $c_1 \ne \pm 1$, $\psi$ satisfies the elliptic equation
\begin{equation}\label{eq38}
{\psi_\xi}^2=\frac{2}{\lambda \gamma}(c_1-\cos \psi)~,
\end{equation}
with solution given by
\begin{equation}\label{eq40}
\psi(\xi)=2 ~\mathrm{am}\left(\mp\sqrt{\frac{c_1-1}{2\lambda\gamma}}(\xi-\xi_0); \frac{2}{1-c_1}\right)~.
\end{equation}
For plots of the sine-Gordon Jacobi amplitude solutions, see Figure~(\ref{m7}). Only the Jacobi amplitude solutions of
superunitary modulus $|m|$ are periodic and bounded and are physically relevant.

\subsection{The sinh-Gordon equation}
The hyperbolic version of the sine-Gordon equation, i.e.
the sinh-Gordon equation, is also extensively used in integrable quantum field theory \cite{sinh1,sinh2}, kink dynamics \cite{sinh3}, and hydrodynamics \cite{sinh4}.
Here, we will use the parametrization in \cite{sinh1} and write the corresponding equation in the variable $h$ in the form
\begin{equation}\label{em33}
\frac{\partial^2}{\partial u \partial v}\log h=\frac{1}{2}\left(h^2-h^{-2}\right)=\sinh(2\log h)~.
\end{equation}
Thus, we identify the constants to be $\alpha=\frac{1}{2},~\beta=-\frac{1}{2},~a=2,~ b=-2$, which gives the quadrature
\begin{equation}\label{em34}
\int \frac{dh}{h\sqrt{c_1+\frac 1 2\cosh (2\log h)}}=\pm\sqrt{\frac{2}{\lambda \gamma}} (\xi-\xi_0)~,
\end{equation}
which is equivalent to
\begin{equation}\label{em35}
\int \frac{d \psi}{\sqrt{c_1+\frac 1 2 \cosh (2\psi)}}=\pm\sqrt{\frac{2}{\lambda \gamma}} (\xi-\xi_0)~.
\end{equation}

In implicit form, $h$ satisfies~(\ref{c1zero}), which simplifies to
\begin{equation}\label{song}
h\, _2F_1\left(\frac 1 2 ,\frac 1 4, \frac 5 4 ;-h^{4}\right)=\pm \frac{\xi-\xi_0}{\sqrt{2\lambda \gamma}}~.
\end{equation}
By a simple transformation, the explicit solution for $c_1=0$ is obtained by solving the integral
\begin{equation}\label{em35}
\int \frac{d \psi}{\sqrt{\cosh (2\psi)}}=\pm\sqrt{\frac{1}{\lambda \gamma}} (\xi-\xi_0)~,
\end{equation}
and using (\ref{eq350}) and (\ref{eq352}) we obtain
\begin{equation}\label{eq352}
\psi(\xi)=i~ \mathrm{am}\left(\pm  \sqrt{\frac{1}{-\lambda \gamma}} (\xi-\xi_0);2\right)~,
\end{equation}
which can also be obtained from (\ref{eq400}) for $c_1=0$.

For the special case of $c_1=-\frac 1 2 $,~(\ref{em35}) simplifies to
\begin{equation}\label{em36}
\int \frac{d \psi}{\sinh \psi}= \pm\sqrt{\frac{2}{\lambda \gamma}} (\xi-\xi_0)~.
\end{equation}
The kink-antikink solutions are
\begin{equation}\label{em37}
\psi(\xi)=2~\mathrm{arctanh} \left(e^{\pm\sqrt{\frac{2}{\lambda \gamma}} (\xi-\xi_0)}\right)~.
\end{equation}

\begin{figure}[ht!] 
\centering 
\includegraphics[width=0.45\textwidth]{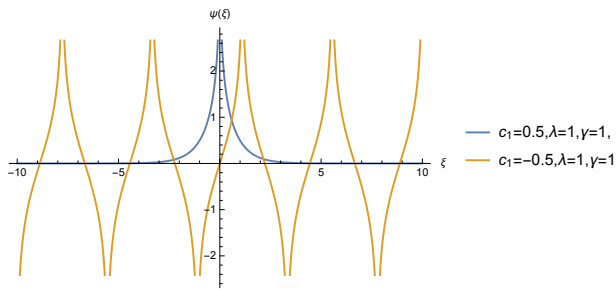}
\caption{The arctanh solutions (\ref{em37}) and (\ref{em41}) of the sinh-Gordon equation.}
\label{m8}
\end{figure}

When $c_1=\frac 1 2$,~(\ref{em35}) simplifies to
\begin{equation}\label{em40}
\int \frac{d \psi}{\cosh \psi}=\mathrm{gd}(\psi)= \pm\sqrt{\frac{2}{\lambda \gamma}} (\xi-\xi_0)~,
\end{equation}
where the Gudermannian function $\mathrm{gd}(\psi)=2~\mathrm{arctan}\left(\tanh \frac \psi 2\right)$
gives also  the kink-antikink solutions
\begin{equation}\label{em41}
\psi(\xi)= 2~\mathrm{arctanh} \left[\tan \left(\pm\frac{1}{\sqrt {2\lambda \gamma}} (\xi-\xi_0)\right) \right]~.
\end{equation}
The arctanh solutions are plotted in Figure~(\ref{m8}).
Being singular, they have only theoretical interest from the point of view of blow-up analysis.

When $c_1 \ne \pm \frac 1 2$, then $\psi$ satisfies the elliptic equation
\begin{equation}\label{em38}
{\psi_\xi}^2=\frac{2}{\lambda \gamma}\left(c_1+\frac 1 2\cosh (2\psi) \right),
\end{equation}
with solutions given by the Jacobi amplitude function
\begin{equation}\label{eq400}
\psi(\xi)= i ~\mathrm{am}\left(\pm\sqrt{\frac{2 c_1+1}{-\lambda\gamma}}( \xi-\xi_0); \frac{2}{2c_1+1}\right)~.
\end{equation}
\begin{figure}[ht!] 
\centering 
\includegraphics[width=0.45\textwidth]{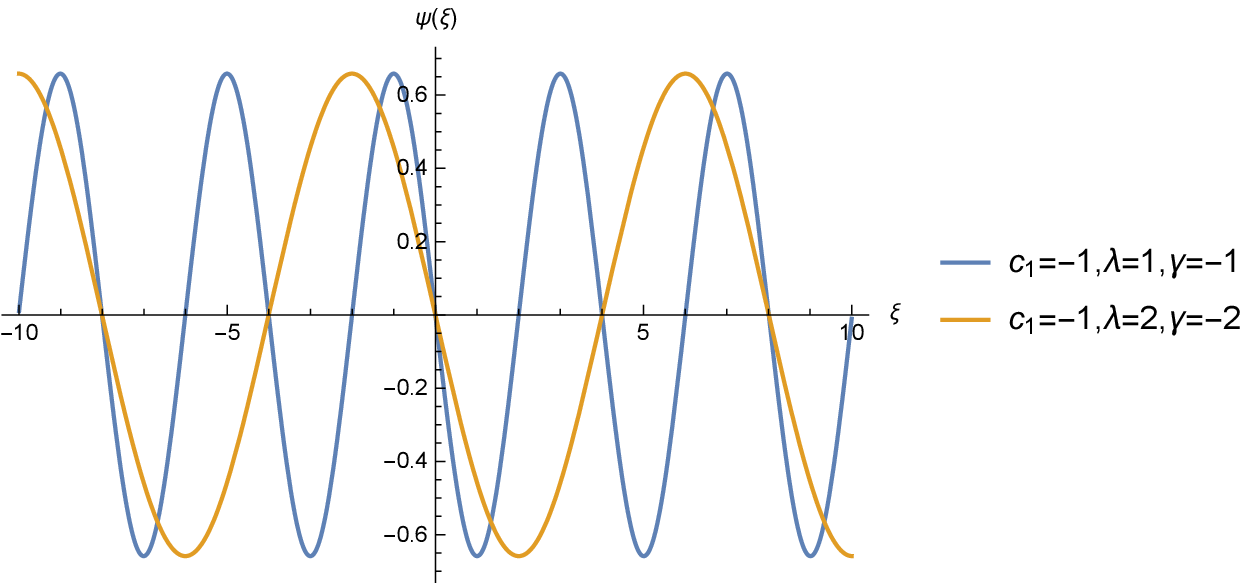}\\
\includegraphics[width=0.45\textwidth]{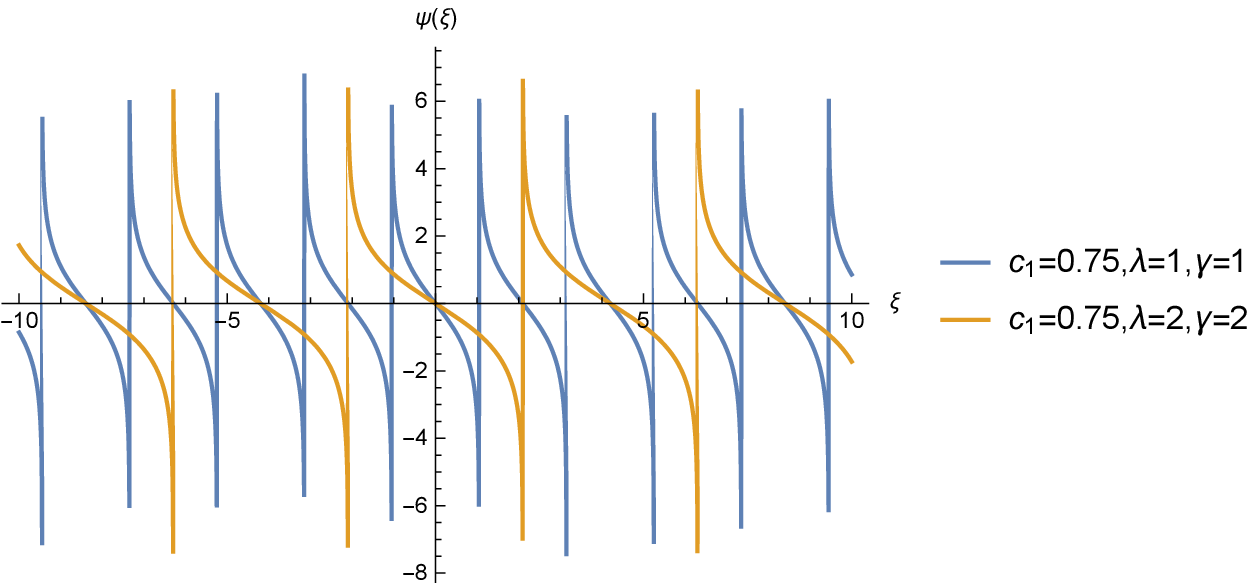}
\caption{The Jacobi amplitude solution (\ref{eq400}) of the sinh-Gordon equation for $|m|>1$ (top) and $|m|<1$ (bottom).}
\label{m9}
\end{figure}
Plots of (\ref{eq400}) are given in Figure~(\ref{m9}). Similarly to the sine-Gordon case, only the Jacobi amplitude solutions of superunitary modulus $|m|$ are
bounded periodic functions.

\newpage

\section{Conclusion}

\medskip

In summary, we have used a very simple method to obtain all the basic soliton, periodic and Weierstrass solutions of wave equations with two exponential nonlinearities whose particular cases correspond to celebrated equations in mathematical physics, such as Liouville, Tzitz\'eica and its variants, sine-Gordon, and sinh-Gordon equations. All these solutions are obtained consistently in the traveling variable by a thorough analysis of elliptic equations. Novel implicit solutions in terms of generic hypergeometric functions are also obtained through a direct integration. Although there are other methods to obtain these translation-invariant solutions, e.g., the integral bifurcation method \cite{rui},
some of these solutions, in particular the Weierstrass solutions of the Tzitz\'eica class of equations and the amplitude Jacobi solutions of the sine/sinh-Gordon equations cannot be obtained by the tanh method usually employed in the literature. Consequently, with a few exceptions in the case of the amplitude Jacobi solutions \cite{rob,mond}, their potential for realistic physical applications has been ignored in the past. As for the the Weierstrass solutions of the Tzitz\'eica class of equations, their potential in the area of optical solitons is still to be
assessed \cite{t-app4, t-app5}.

We also plan a future publication with the aim to extend the unifying approach presented in this paper to the hyperelliptic cases \cite{bf}.

Finally, for more complicated multiple-soliton (multi-phase soliton) solutions, one should use Darboux and B\"acklund transformations \cite{boldin,brejnev,conte}.

\bigskip
\bigskip
\bigskip



{\bf Acknowledgements}

\medskip

SCM would like to acknowledge partial support from the Dean of Research \&
Graduate Studies at Embry-Riddle Aeronautical University while on a short visit to IPICyT,  San Luis Potos\'{\i}, Mexico.
MPM thanks CONACyT for a doctoral fellowship.


\bigskip
\bigskip


%


\end{document}